\documentclass{article} 
\usepackage{nips12submit_e,times}
\usepackage{graphicx}
\usepackage{subfigure}
\usepackage{amsmath}
\usepackage{amssymb}
\usepackage{url}
\usepackage{color}
\usepackage{multirow}
\usepackage{booktabs}

\newcommand{\bs}[1]{\mathbf{#1}}
\newcommand{\bgs}[1]{\boldsymbol{#1}}
\newcommand{\tr}{T}
\newcommand{\eq}[1]{\begin{equation} #1 \end{equation}}

\newcommand{\R}{\mathbb{R}}

\newcommand{\sdx}{z}
\newcommand{\sox}{x}
\newcommand{\sey}{w}
\newcommand{\sdy}{v}
\newcommand{\soy}{y}
\newcommand{\sop}{\theta}
\newcommand{\pn}{\bgs{\epsilon}}    
\newcommand{\pdx}{\bs{\sdx}}
\newcommand{\pox}{\bs{\sox}}
\newcommand{\pey}{\bs{\sey}}
\newcommand{\pdy}{\bs{\sdy}}
\newcommand{\poy}{\bs{\soy}}
\newcommand{\pop}{\bgs{\sop}}
\newcommand{\Mop}{\bgs{\Theta}}

\newcommand{\std}{\eta}

\author{
Sebastian Bitzer, Izzet B.~Yildiz and Stefan J.~Kiebel\\
Max Planck Institute for Human Cognitive and Brain Sciences, Leipzig\\
\texttt{\{bitzer,yildiz,kiebel\}@cbs.mpg.de}
}

\nipsfinalcopy 

\begin{document}
\title{Online Discrimination of Nonlinear Dynamics\\ with Switching Differential Equations}
\maketitle

\begin{abstract}
How to recognise whether an observed person walks or runs? We consider a dynamic environment where observations (e.g. the posture of a person) are caused by different dynamic processes (walking or running) which are active one at a time and which may transition from one to another at any time. For this setup, switching dynamic models have been suggested previously, mostly, for linear and nonlinear dynamics in discrete time. Motivated by basic principles of computations in the brain (dynamic, internal models) we suggest a model for switching nonlinear differential equations. The switching process in the model is implemented by a Hopfield network and we use parametric dynamic movement primitives to represent arbitrary rhythmic motions. The model generates observed dynamics by linearly interpolating the primitives weighted by the switching variables and it is constructed such that standard filtering algorithms can be applied. In two experiments with synthetic planar motion and a human motion capture data set we show that inference with the unscented Kalman filter can successfully discriminate several dynamic processes online. 
\end{abstract}

\section{Introduction}
Humans are extremely accurate when discriminating rapidly between different dynamic processes occurring in their environment. For example, for us it is a simple task to recognise whether an observed person is walking or running and we can use subtle cues in the structural and dynamic pattern of an observed movement to identify emotional state, gender and intent of a person. It has been suggested that this remarkable perceptual performance is based on a learnt, generative model of the dynamic processes in the environment which the brain uses to infer the current state of the environment given sensory input \cite{Doya2007}. Motivated by this view we propose a generative model based on continuous-time dynamics which models different dynamic processes in the environment by switching between different nonlinear differential equations. Using online inference in this model we can rapidly and accurately discriminate between different dynamic processes, e.g., motions. 

In the model, a Hopfield network \cite{Hopfield1984} implements the dynamics between switching variables (the "switching dynamics"). The Hopfield dynamics implements a winner-take-all mechanism between arbitrary many switching variables such that only one of the switching variables is active in each stable fixed point of the dynamics. We associate each switching variable with different parameters of a parametric differential equation implemented by a dynamic movement primitive (DMP) \cite{Ijspeert2003,Schaal2007}. The parameters are then interpolated based on the continuous values of the switching variables and the resulting differential equation (the observation dynamics) is used to generate observations.

Exact inference in the nonlinear, continuous-time, hierarchical model is intractable. Here, we show that a standard filtering procedure (the unscented Kalman filter) enables efficient, robust and, importantly, online discrimination of dynamic processes. We illustrate these features using experiments with (i) simple nonlinear movements in a plane and (ii) motion capture data of a human walking in different styles.

\subsection{Related Work}
Switching dynamic models are well-established in statistics \cite{Fruhwirth-Schnatter2006,Chen2000}, signal processing \cite{Chen2003} and machine learning \cite{Ghahramani2000,Murphy2002,Quinn2009}. In contrast to these models, we define both, the dynamic models and the switching variables, using nonlinear dynamical systems with continuous states running in continuous time.  This allows us to link our model more easily to computations implemented in analogue biological substrate such as the brain. Additionally, a formulation in continuous time allows us to easily perform time-rescaling of dynamical systems \cite{Ijspeert2003}.

More recently, other continuous-time switched dynamic models have been proposed, for example, nonparametric models \cite{Garnett2010,Saatci2010} which extend Bayesian online change point detection \cite{Adams2007,Fearnhead2007} using Gaussian processes. Although online inference methods for these models have been described, their aim is not to identify a known dynamic process, but rather to make accurate predictions of observations across change points at which the underlying dynamic process changes. Similarly, switched latent force models \cite{Alvarez2010} are nonparametric models in which the position of change points and the underlying dynamic processes are modelled  using Gaussian processes and DMPs. The proposed inference method is offline, i.e., all observed data are used and again the aim is not to discriminate between different, previously learnt models. Rather, this approach could be used to learn parametric models based on the obtained change point posterior.

In \cite{Opper2010}, the authors derive a smoothing algorithm based on variational inference for an Ornstein-Uhlenbeck (OU) process which is switched by a random telegraph process. Thus, this model can only switch between two constant drifts. Similarly, \cite{Stimberg2011,Stimberg2012} propose Markov chain Monte Carlo inference for a switched OU process where the number of different parameter sets as well as the parameter values are automatically determined from the data. However, these parameters are limited to the constant drift and diffusion parameters of the OU process which cannot implement generic nonlinear dynamics. In contrast to these models, we approximate a change point process using our continuous-valued switching dynamics which allows us to maintain a coherent continuous framework and apply standard filtering algorithms.

In the following we describe the present model and the proposed inference method in detail and subsequently present results of our experiments in Section \ref{sec:results}.

\section{Switching Dynamics}
\label{sec:switchingdyn}
We want the switching dynamics to be able to form a stable representation of the identity of a dynamical system. We implement this requirement with a Hopfield network \cite{Hopfield1984} which defines winner-take-all dynamics using lateral inhibition between units in a fully connected network. In particular, the network is defined as
\eq{\label{eq:discdyn}
    \dot{\pdx} = k\left( \bs{M}\bgs{\sigma}(\pdx) + b^\mathrm{lin}(g\bs{1} - \pdx) \right) + \pey
}
where $\pdx \in \R^{N}$ are the state variables (one for each of the $N$ dynamical systems in the observation dynamics), $k$ is a rate constant, $\bgs{\sigma}$ is a multidimensional logistic sigmoid function, $b^\mathrm{lin}$ is a parameter determining the strength with which the states tend to converge to the goal $g$, $\bs{1}\in \R^{N}$ is a vector of 1s, $\bs{M}\in \R^{N\times N}$ is a connection matrix and $\pey \in \R^{N}$ is external input to the dynamical system. Note that we will use these external inputs $\pey$ to induce the actual switching behaviour (see below). Lateral inhibition for winner-take-all dynamics is implemented using, 
\eq{\label{eq:sigmoid}
    \sigma_i(\pdx) = \sigma(\sdx_i) = \frac{1}{1 + \exp(-r(\sdx_i - o))} \qquad \mathrm{and} \qquad \bs{M} = b^\mathrm{lat} (-\bs{1} + \bs{I})
}
where now $r$ determines the slope and $o$ the centre of $\sigma$, $b^\mathrm{lat}$ is a parameter determining the strength of lateral inhibition, $\bs{1}\in \R^{N\times N}$ is a matrix of 1s and $\bs{I}$ is the identity matrix. The dynamics in eq. \eqref{eq:discdyn} has two parts: 1) the lateral inhibition $\bs{M}\bgs{\sigma}(\pdx)$ and 2) the linear term $b^\mathrm{lin}(g\bs{1} - \pdx)$ which attracts the state $\pdx$ to $g\bs{1}$. Note that the fixed points of the system are implicitly given as
\eq{
    \sdx_i = g - \frac{b^\mathrm{lat}}{b^\mathrm{lin}} \sum_{j\neq i} \sigma(\sdx_j) \quad \forall i\in 1, \dots, N.
}
One can see that the fixed points with one state $\sdx_m \approx g$ while all others are $\sdx_{j\neq m} \approx 0$ are local minima of the underlying Lyapunov function and therefore stable \cite{Hopfield1984} provided that $o=g$ and $b^\mathrm{lat}/b^\mathrm{lin} = 2g$. In the experiments below we chose $g = 10$, $b^\mathrm{lat} = 1.7$, $k=4$, $r=1$ and the remaining parameters according to the given constraints.

We define the output of the switching dynamics as 
\eq{
    \pdy = \bgs{\sigma}^\sdy(\pdx)
}
such that the elements of $\pdy \in \R^{N}$ lie between 0 and 1. The output $\pdy$ is then used as the weights of a linear interpolation of the observation dynamics parameters: $\pop = \Mop\pdy$, where $\pop\in\R^{M}$ are the parameters currently used at the observation level and $\Mop\in\R^{M\times N}$ are the $N$ sets of parameters of the $N$ dynamical systems in the observation dynamics. By setting the limits of $\pdy$ to 0 and 1 we, therefore, ensure that only one set of parameters is active for each stable fixed point of the switching dynamics. We here use the logistic sigmoid function (eq. \ref{eq:sigmoid}) with $o = g/2$ and $r = 0.7$. We chose $r=0.7$ to increase the region within the interval $[0,g]$ for which the logistic function is approximately linear while still maintaining $\sigma(0)\approx 0$ and $\sigma(g)\approx 1$. This setting eases inference of $\pdx$ later on.

Evolving the switching dynamics leads to the following behaviour: $\pdx$ converges to the fixed point $\sdx_i\approx g, \sdx_{i\neq j}\approx 0$ (fixed point $i$) where $i$ is the index of the state which was the largest at $t=0$. In the experiments below, when generating data, we induce switching by applying an external input $\sey_k = 4$ for a short period of time in order to switch from fixed point $i$ to fixed point $k$.

\section{Representing Dynamical Systems with Dynamic Movement Primitives}
\label{sec:DMPs}
In the observation dynamics we use a simplified form of rhythmic dynamic movement primitives (DMPs) \cite{Ijspeert2003,Schaal2007} to represent rhythmic dynamical systems. By using DMPs we gain the following properties:
\begin{enumerate}
    \item arbitrary, rhythmic motion is easily learnt
    \item different DMPs describing different movements have a common parameterisation
    \item learnt dynamical systems represent a strongly attracting limit cycle which improves discriminability during inference
\end{enumerate}
We define a DMP as
\eq{\label{eq:obsdyn}
    \begin{split}
            \dot{\omega} &= f_\omega(\pdy) \\
            \dot{\bs{s}} &= a\left(\bs{f}(\omega,\pdy) - \bs{s}\right)
    \end{split} \qquad
    \bs{f}(\omega,\pdy) = \frac{\bs{k}(\omega)^\tr}{\bs{1}^\tr\bs{k}(\omega)}\bs{W}(\pdy)
}
with phase variable $\omega$ and position variables $\bs{s}$. The phase $\omega$ is governed by a constant drift which implements the motion around the limit cycle. Its frequency $f_\omega(\pdy)/(2\pi)$ depends on the switching dynamics through $\pdy$ (see Section \ref{sec:switchingdyn}). The position variables $\bs{s}\in\R^D$ describe the actual motion through $D$-dimensional space. For fixed phase they implement a simple linear point attractor at $\bs{f}(\omega,\pdy)$ where $a>0$ determines the strength of attraction. Thus, for drifting phase the position variables follow $\bs{f}(\omega,\pdy)$ such that the complete dynamical system implements a limit cycle at the positions defined by $\bs{f}(\omega,\pdy)$. We choose $\bs{f}(\omega,\pdy)$ to be a linear combination of (normalised) basis functions $\bs{k}(\omega)$. In particular, we use circular von Mises basis functions as in \cite{Schaal2007}: $ k_j(\omega) = \exp\left( (\cos(\omega - c_j) - 1)/\ell^2 \right)$ with width $\ell>0$ which reach their maximum value of 1 at centres $c_j$.

In the experiments below we fixed $a=1$, which is sufficiently large such that the states quickly return to the limit cycle when perturbed, but other values work similarly well. We determined $c_j$ and $\ell$ from the number of basis functions $C$ as follows: We evenly distributed the centres $c_j$ in $[0,2\pi]$ and set $\ell=2\pi/C$. We chose the frequency $f_\omega$ for each data set by an ad-hoc estimate of the cycle period described by the data. More sophisticated methods are available \cite{Gams2009}, but were not required for the demonstration considered here. Finally, we learnt the weights of the basis functions $\bs{W}$ from given positions $\pox(t_i), i=1,\dots,T$ by computing $\bs{k}(\omega)$ from the phases $\omega(t_i)$\footnote{By default we set $\omega(t_1)=0$ for all data sets.} and applied simple least squares fitting to the resulting data set as suggested in \cite{Ijspeert2003,Schaal2007}.

Furthermore, we allow for an affine transformation of the position variables which produces the actual observations of the model: $\poy = \bs{A}\bs{s} + \bs{b}$.

\section{Online Inference for Switching Dynamical Systems}
The dynamical systems described in Sections \ref{sec:switchingdyn} and \ref{sec:DMPs} are the basis for a hierarchically structured generative model. We combine all dynamic states into a single vector $\pox$ such that $\dot{\pox} = [\dot{\pey}^\tr, \dot{\pdx}^\tr, \dot{\omega}, \dot{\bs{s}}^\tr]^\tr$. By assuming Gaussian noise on the variables in the model we obtain the following stochastic differential equation
\eq{\label{eq:filterproblem}\begin{split}
    d\pox &= \left[ \begin{array}{c}
            -\pey\\
            k\left( \bs{M}\bgs{\sigma}(\pdx) + b^\mathrm{lin}(g\bs{1} - \pdx) \right) + \pey\\
            f_\omega(\pdy) \\
            a\left(\bs{f}(\omega,\pdy) - \bs{s}\right)
    \end{array} \right]dt + \bs{B}d\pn_\sox\\
    \poy &= \bs{A}\bs{s} + \bs{b} + \pn_\soy
\end{split}}
where $\pdy = \bgs{\sigma}^\sdy(\pdx)$, $d\pn_\sox$ represents a Wiener process, $\bs{B} = \mathrm{diag}(\bgs{\std})$ with $\bgs{\std} = [\bgs{\std}_\sey^\tr,\bgs{\std}_\sdx^\tr,\std_\omega, \bgs{\std}_s^\tr]^\tr $ being the standard deviations for each variable such that $\bs{Q} = \bs{B}\bs{B}^\tr$ is the corresponding covariance matrix. The Gaussian measurement noise $\pn_\soy$ has covariance matrix $\bs{R} = $diag$(\bgs{\std}_\soy^2)$. We model external inputs to the Hopfield network $\pey$ with an Ornstein-Uhlenbeck process which fluctuates around $\bs{0}$. Although this process does not generate brief, switch-inducing inputs (as e.g. expected when switching gaits), we will show below that it provides sufficient flexibility such that switches can be reliably inferred from the data.

Our aim is to infer the state $\pdx(t)$ of the switching dynamics based on the history of observed data $\poy(\tau), \tau\leq t$ which is input to the observation dynamics. In general, this is a filtering problem which could be solved optimally with the Kalman-Bucy-Filter \cite{Jazwinski1970}, if the dynamical systems were linear. For our nonlinear dynamical systems we have to resort to approximate filtering methods such as the extended \cite{Jazwinski1970} or unscented Kalman filters (UKF) \cite{Julier1995,Wan2001a}, or particle filters \cite{Doucet2001}. Here, we used the UKF because of its balanced trade-off between computational efficiency and nonlinear filtering performance. While a continuous-time version of the UKF has been suggested \cite{Sarkka2007} we found this to be numerically unstable for the present filtering problem and instead we used the standard discrete UKF with the following approximation in the prediction step: we numerically integrated the deterministic part of eq. \eqref{eq:filterproblem} starting from the current sigma points of the UKF to approximate the nonlinearly transformed sigma points and then added $\Delta t\bs{Q}$ to the estimated posterior covariance instead of $\bs{Q}$ (cf. eq. 19 in \cite{Sarkka2007}). Here, $\Delta t$ is the time between the last and the new data point and the factor of $\Delta t$ corresponds to the variance of noise which, prescribed by the Wiener process, has been accumulating within $\Delta t$. We used standard settings of UKF parameters as reported in \cite{Wan2001a}, i.e., we chose $\alpha=0.01$, $\beta=2$ and $\kappa=3-L$ where $L$ is the dimensionality of $\pox$, $L = 2N + D + 1$.

The UKF requires prior choices for the covariance matrices of the states and observations, $\bs{Q}$ and $\bs{R}$. Below we used the same $\bs{Q}$ for inference on both data sets where we set $\bgs{\std}_\sey = 1, \bgs{\std}_\sdx = 0.5, \std_\omega = 0.05$ and $\bgs{\std}_s = 0.1$ (the scalars are expanded to all dimensions of the corresponding vectors). These values have been manually selected to facilitate switching during inference and embody a high level of prior uncertainty about the hidden states. Smaller values for $\bgs{\std}$ may be chosen (up to two orders of magnitude) with only moderate loss in discrimination performance. However, we found that $\bgs{\std}_\sey$ should be kept high as otherwise the UKF will be more likely to miss a switch. The choice for $\bs{R}$ is experiment-specific and is described below.
\section{Experiments}
\label{sec:results}
To provide a proof of concept, we modelled (i) synthetic data of planar motion and (ii) human motion capture data. The experiments below show that UKF filtering can discriminate between different motions. To measure performance we filtered a long stream of data consisting of a large number of trials between which the observed motion switched. The goal was to identify, as fast as possible, the DMP that caused the data, i.e., to infer switches between DMPs accurately. We report the fraction of correct model responses (or \% correct of the total number of trials) where we define the system response as the identity of the maximal switching variable after a chosen number of time steps from the beginning of a trial: $\arg\max_i \sdx_i(n\times\Delta t)$.

\begin{figure}[p]
  \begin{center}
   \subfigure[noise-free trajectories]{\label{fig:4choiceDDot}
         \includegraphics[width=.3\linewidth]{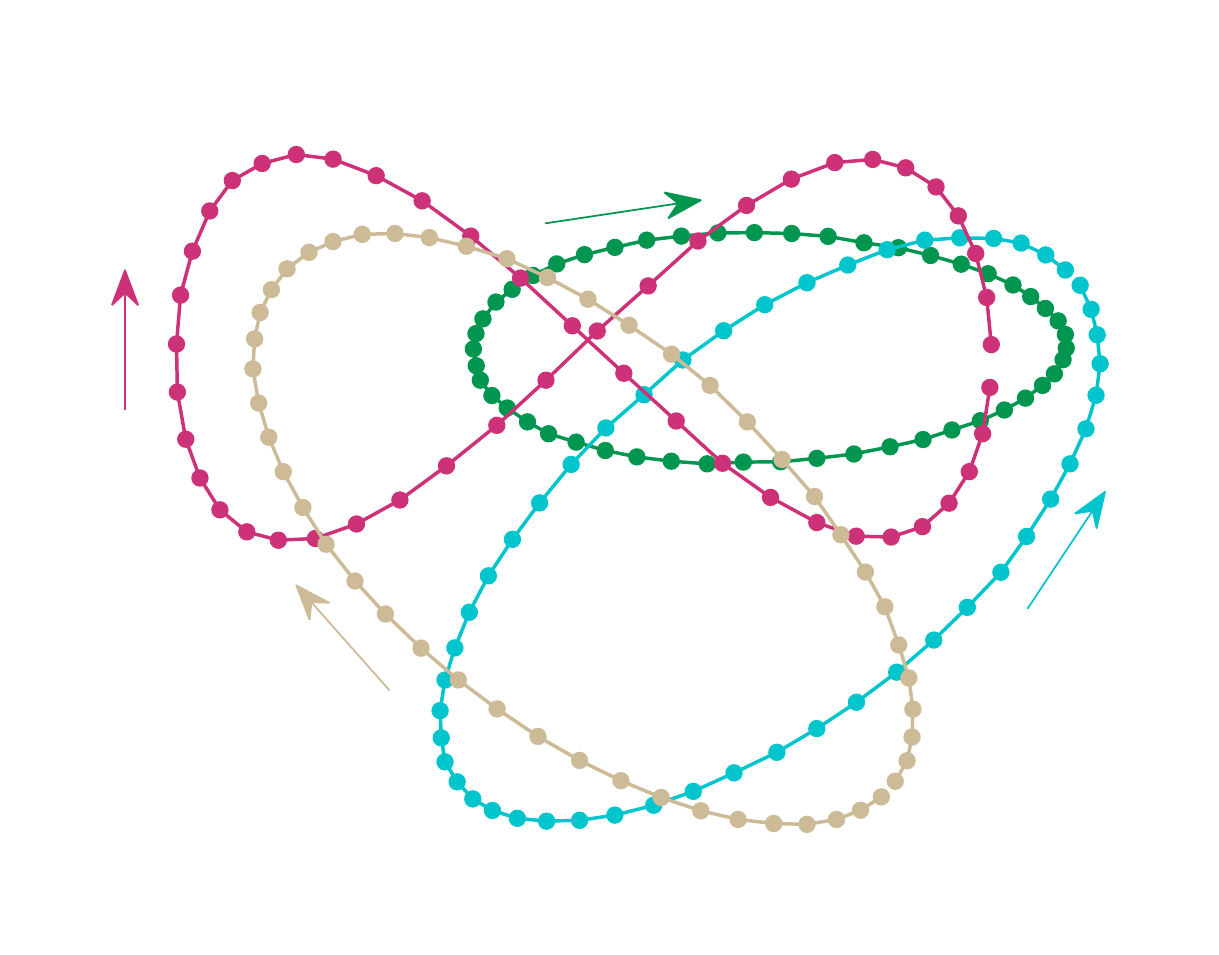}
    }
    \subfigure[noisy trajectories]{\label{fig:ellnoisetraj}
        \includegraphics[width=.3\linewidth]{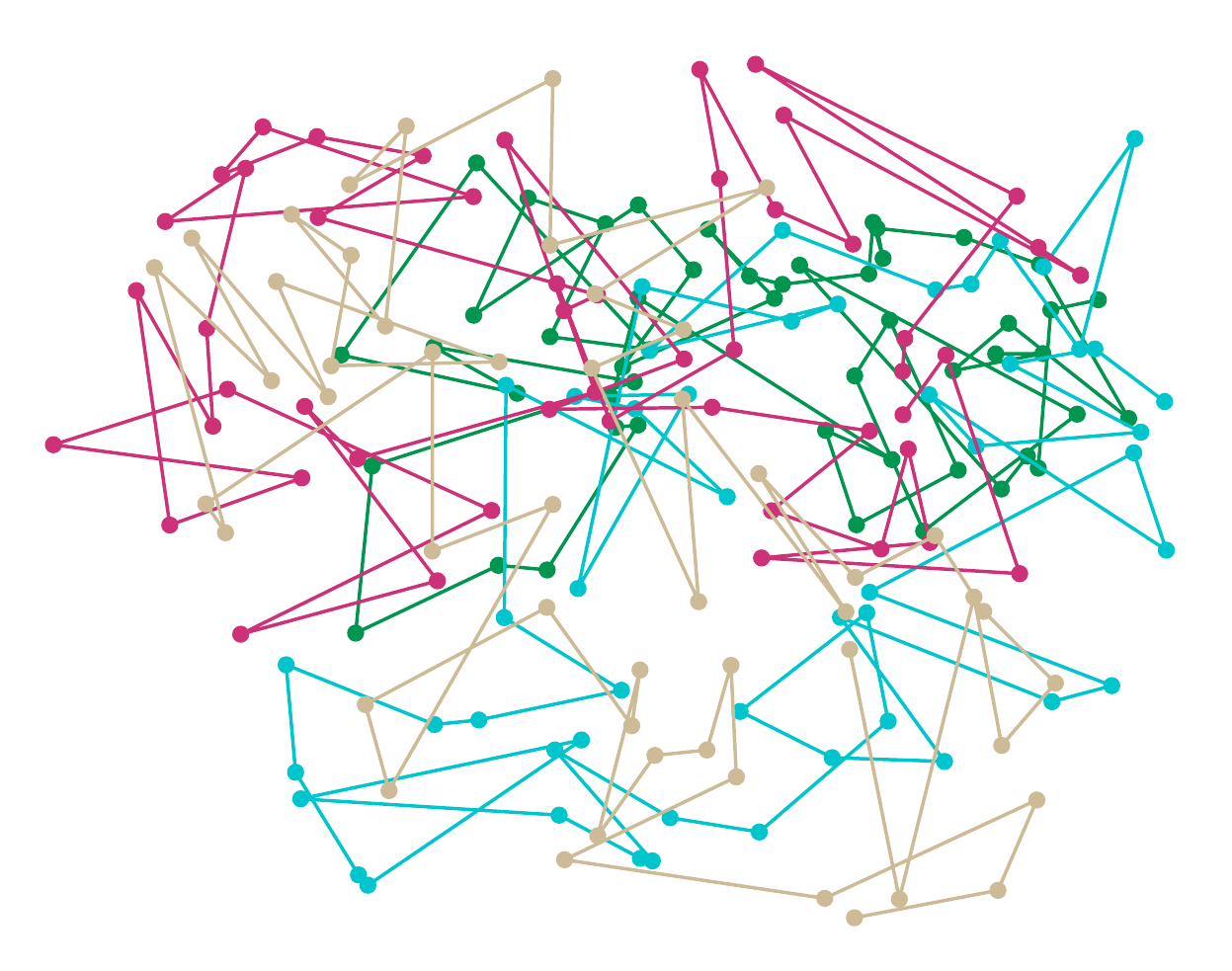}
    }
    \subfigure[original]{\label{fig:walker}
         \includegraphics[width=.12\linewidth]{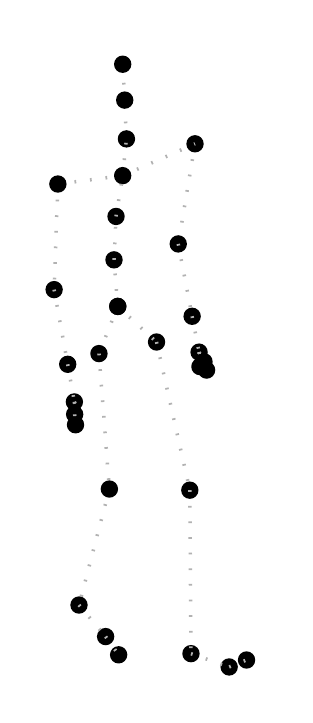}
    }
    \subfigure[noisy]{\label{fig:walkernoise}
         \includegraphics[width=.12\linewidth]{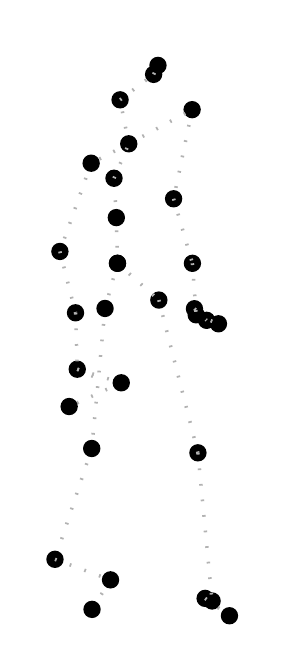}
    }
  \end{center}
  \caption{Illustration of the used motions. (a) Trajectories on which the two observed dots move for the four synthetic planar motions. Motion direction indicated by arrows. (b) One example period of each motion of (a) perturbed by observation noise with $\bgs{\eta}_\soy=0.2$. (c) A typical motion capture posture (one frame of a walk). Observed data are the 3D positions of the dots (connecting lines shown for illustration purposes only). (d) Posture shown in (c) perturbed by noise as used in our experiment.
} \label{fig:data}
\end{figure}

\subsection{2D dots}
\begin{figure}[p]
  \begin{center}
   \subfigure[inference on noise-free data]{\label{fig:ellresults}
         \includegraphics[width=\linewidth]{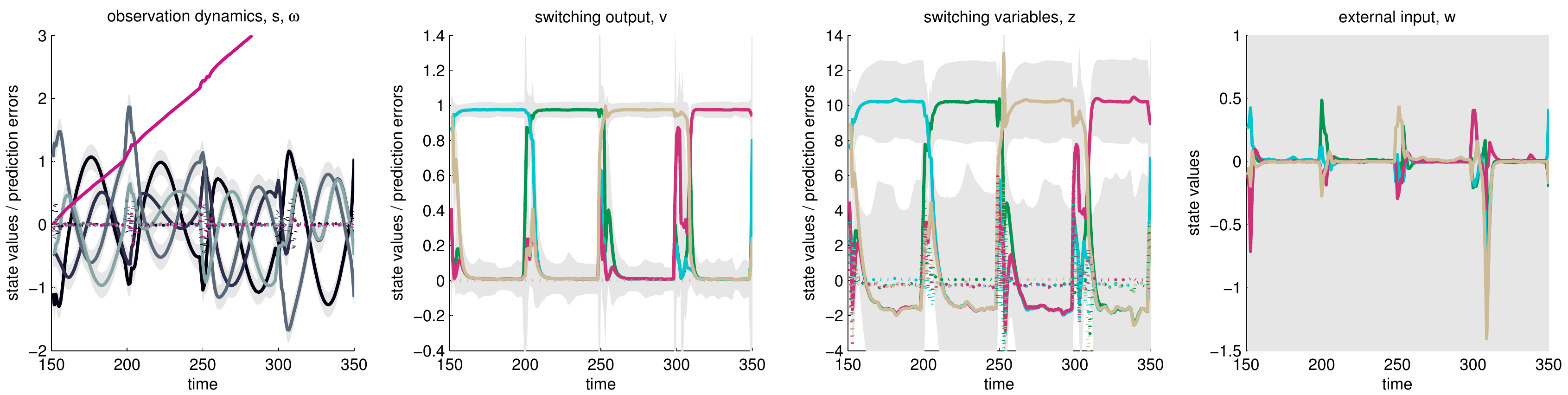}
    }\\
    \subfigure[inference on noisy data]{\label{fig:ellnoiseresults}
         \includegraphics[width=\linewidth]{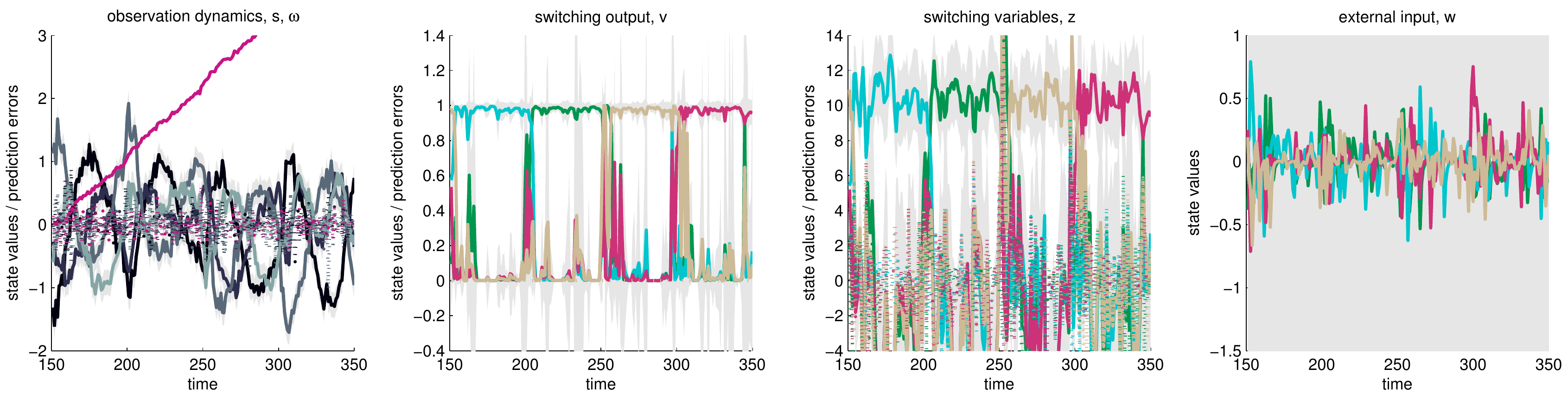}
    }
  \end{center}
  \caption{Discriminating four planar motions. (a) Inference results for four example trials on noise-free data. The observed motion switches every 50 time steps. Shown are (from left to right): (i) the DMP positions $\bs{s}$ (grey) and phase $\omega/(2\pi)$ (red, rising), (ii) the switching output $\pdy$, (iii) switching variables $\pdx$, and (iv) the external input $\pey$. Dotted lines show the prediction errors (difference between predicted states and inferred states after UKF update). Grey shading indicates two times the posterior standard deviation of the corresponding variables (for switching output determined via unscented transform from switching variables). The true sequence of movements in terms of colours is light blue-green-yellow-red which is found by the model after a short transient period at the beginning of each trial (as determined by finding the switching variable with the highest value). (b) Same as (a), but for noisy movements. The inferred states become more noisy, but movements can still be reliably discriminated.
} \label{fig:ellipses}
\end{figure}

We first illustrate inference in our model using a simple example of synthetic motions of dots which followed different trajectories on a plane. The chosen trajectories formed three different ellipses and a figure-8 and were partially overlapping as depicted in Fig. \ref{fig:4choiceDDot}. The observations were the 2D-positions of two phase-shifted (by $\pi$) dots. Thus, the online inference task was to determine by which DMP the motion of the two dots was generated. The experimental procedure was as follows: 1) learn the four DMPs from example data, 2) generate dynamically switched data from learnt DMPs (25 trials for each DMP) and 3) infer hidden variables from generated data.

We set the size of time steps and frequency of oscillation arbitrarily such that a full period of the oscillation was reached after 50 time steps. We used $C=7$ basis functions and the identity function as output function, i.e., $\bs{A} = \bs{I}$ and $\bs{b} = \bs{0}$, i.e. the dots directly plot the on-going dynamics. Hence, the number of observations was equal to the dimensionality of the position variables $\bs{s}$, $D=4$.

For generating data from the switching model we set all prior uncertainties to $0.001$ (negligibly low noise) unless specified otherwise below. We then generated 25 trials for each of the four motions in the data set by simulating $50\cdot 25\cdot 4$ time steps while randomly switching to a different than the current fixed point attractor every 50 time steps using the external input to the Hopfield dynamics $\pey$.

For online inference based on eq. \eqref{eq:filterproblem} and the generated stream of data we initialised the dynamic variables in the model with random values. In particular, we started the switching dynamics randomly in one of the fixed points, set the position variables $\bs{s}$ in the observation dynamics to the 0-position on the trajectory of the chosen attractor and chose the phase variable $\omega$ randomly (uniformly) in the interval $[-\pi/4,\pi/4]$. Note that the initialisation is only relevant for the very first trial of each experiment as we define the beginning of all other trials as the time point of a switching impulse in $\pey$.  Further, we set the prior uncertainty of the observations to the true value $\bgs{\std}_\soy = 0.001$.

For these data the model quickly switched into the correct attractor of the switching dynamics in all trials resulting in 100\% correct responses after less than 15 observations into each trial, i.e., each DMP was correctly identified after less than observing 30\% of a cycle  (cf. Table \ref{tab:fracCorrect} and Fig. \ref{fig:ellresults} for a typical example). 

We tested the robustness against noise by generating movement sequences, but with increased observation noise $\bgs{\std}_\soy$. For our choice of $\bgs{\std}_\soy = 0.2$ the random steps in the plane introduced by noise were on average more than twice as big as the steps introduced by the movements themselves, i.e., at any single point in time random velocities masked the velocities prescribed by the movements (see Fig. \ref{fig:ellnoisetraj}). Nevertheless, dynamic inference still performed at almost 100\% correct responses but needed longer into a movement cycle to gain high performance, as compared to the noise-free case. Even when the prior uncertainty was set to an inappropriately low value ($\bgs{\std}_\soy = 0.01$), the performance stayed above 90\%  (cf. Table \ref{tab:fracCorrect}) after having observed half the cycle.

\begin{table}
    \small
    \centering
    \caption{All experiments: Fraction of correct responses as function of observation time. For 2D ellipses $\Delta t$ is arbitrarily chosen. For walks $\Delta t$ corresponds to 33ms in real time. Even for misspecified models (observation noise in model, $\bgs{\std}_\soy=0.01$, up to two orders of magnitudes smaller than in data), the fraction of correct responses remains high for noisy input.}
    \label{tab:fracCorrect}
    \begin{tabular}{c r r r r r r r r r r}
\toprule
 & $\times \Delta t$ & 5 & 15 & 25 & 35 & 45 & 55 & 65 & 75 & 85\\
\midrule
\multirow{4}{*}{true $\bgs{\std}_\soy$} & ellipses & $0.91$ & $1.00$ & $1.00$ & $1.00$ & $1.00$ &  &  &  & \\
& ell noise & $0.56$ & $0.91$ & $0.98$ & $1.00$ & $0.99$ &  &  &  & \\
& walks & $0.48$ & $0.74$ & $0.74$ & $0.88$ & $0.92$ & $0.94$ & $1.00$ & $1.00$ & $1.00$\\
& walks noise & $0.50$ & $0.66$ & $0.72$ & $0.78$ & $0.86$ & $0.86$ & $0.88$ & $0.90$ & $0.92$\\
\midrule
\multirow{2}{*}{$\bgs{\std}_\soy = 0.01$} & ell noise & $0.51$ & $0.84$ & $0.92$ & $0.98$ & $0.93$ &  &  &  & \\
& walks noise & $0.44$ & $0.70$ & $0.70$ & $0.72$ & $0.76$ & $0.74$ & $0.84$ & $0.80$ & $0.82$\\
\bottomrule
\end{tabular} 
\end{table}

\subsection{Human Walks}
\begin{figure}[t]
  \begin{center}
    \subfigure[inference on original data]{\label{fig:walksresultsnonoise}
        \includegraphics[width=\linewidth]{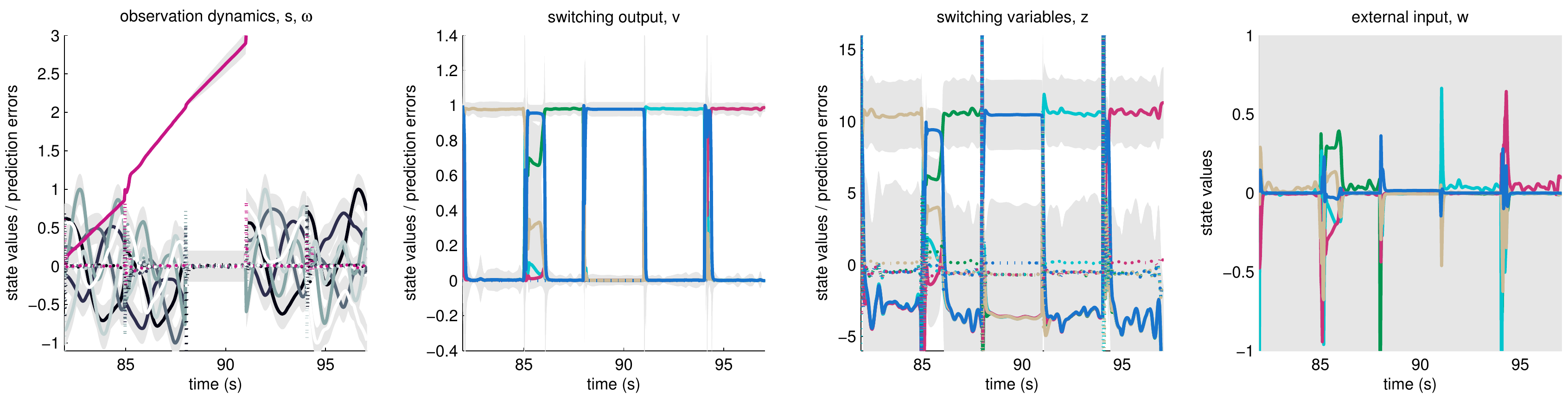}
    }\\
    \subfigure[inference on noisy data]{\label{fig:walksresultsnoisy}
        \includegraphics[width=\linewidth]{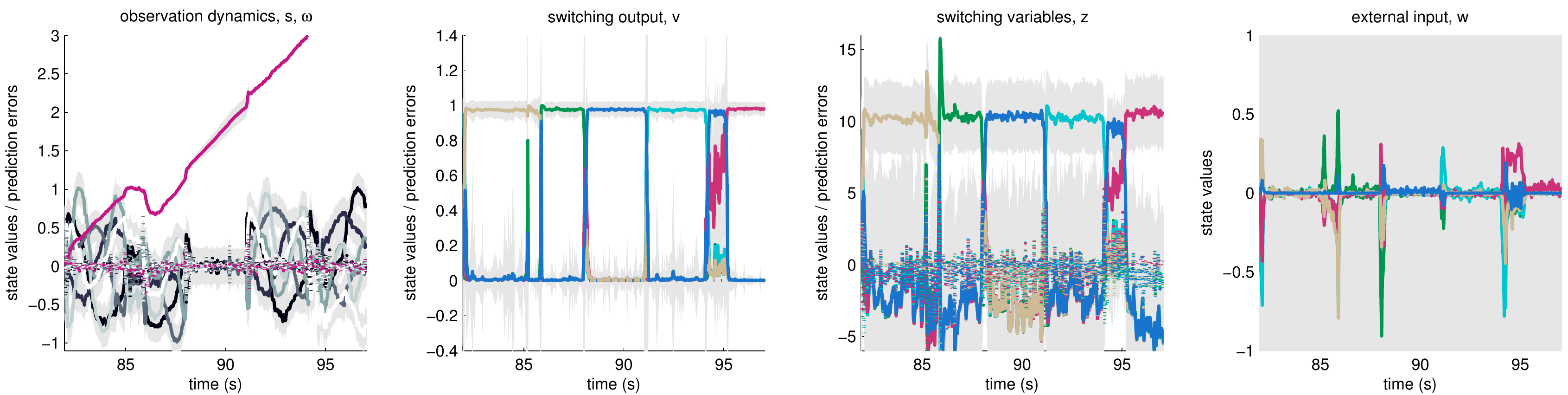}
    }
  \end{center}
  \caption{Inference results for five example trials of the motion capture walks. Format of the figure as in Fig. \ref{fig:ellipses} (a,b). (a) Results for original, concatenated motion capture data. (b) Results for same data with added independent noise. In all five trials the model identified the correct walk (yellow - green - dark blue - light blue - red). From seconds 85 and 94 in (a,b) the model transiently found an alternative explanation for the data (e.g. for second 85 in (a) a linear combination of DMPs), but it switched into the correct attractor of the switching dynamics after at most ca. 1s of real time (ca. 30 motion capture frames) in all shown examples.
}
  \label{fig:walksresults}
\end{figure}

The second experiment is aimed at showing that the present model can in principle also be applied to complicated real-world motions. For this purpose we chose a motion capture data set of a human walking in four different, but similar styles\footnote{motions 1, 5, 15 and 19 of subject 142 of the CMU motion capture database (\url{http://mocap.cs.cmu.edu/}) corresponding to childish, depressed, sad and shy walks, respectively}. As our focus was on the switching model rather than on presenting a complete model for motion capture data, we preprocessed the walks in the following way: To focus on the important aspects of the walk dynamics we removed the global translation of the body. We then computed the position of each motion capture marker in 3D space resulting in a data set of 30 dots moving in 3D space (see Fig. \ref{fig:walker}). We selected a subset of captured frames which roughly contained one period of each walk and which spanned 4s of real time. Because the dynamics of the 30 dots were highly correlated across walks, we performed principle component analysis (PCA) on the data of all walks such that all walks were represented in a common 6D space (capturing ca. 98\% of the original variance) and normalised the data in this space to 0 mean and maximum 1 in each dimension. We estimated the period of each walk by estimating when the trajectory of a walk came closest to finishing a cycle in the normalised space, resulting in 3.17s, 3.8s, 3.03s and 4s for the four walks. We used these periods to estimate the $f_\omega$ and we learnt four corresponding DMPs in the normalised PCA space (using $C=12$ basis functions). In addition, we introduced another, trivial DMP with $\bs{W}=\bs{0}$ to represent constant input which is used to infer the absence of motion, e.g., when the walker stands still.

We integrated these five DMPs into the generative model, where the resulting number of switching variables $\pdx$ was $N=5$ and the number of position variables $\bs{s}$ was $D=6$. Further, we set the linear output transformation of the observation layer ($\bs{A}$ and $\bs{b}$) such that it implemented the mapping from normalised PCA space to the original 3D marker positions. Therefore, the observations in the  model were the 90-dimensional marker position vectors.

To render the inference performed by the system more challenging, we tested the model on the original data (as opposed to data generated by the generative model) where we switched between the four walks (and a fifth standstill walker to test switching to the trivial DMP) every 3s. This harsh switching regime introduced jumps in the underlying phase of the walking cycle in addition to the transition to a different walk. These data had a total length of 150s, consisting of 50 3s walks.

For inference we set the prior uncertainty of the observations to a low $\bgs{\std}_\soy=0.01$ and
used the same initialisation as in the first experiment. Example trials are shown in Fig. \ref{fig:walksresults}. Despite the increase in complexity of the data (higher dimensionality, more dynamical systems), the model switched to the true DMP within the first second in most cases and overall achieved 100\% correct responses (cf. Table \ref{tab:fracCorrect}). 

In addition, the inference is robust against noise: We added Gaussian noise directly on the marker positions with standard deviation equal to that of the marker positions ($\std_\soy \in [2\cdot 10^{-5}, 8.53]$) which, in velocities, translated to 10 times larger noise than signal. Yet, with $\bgs{\std}_\soy$ set to the noise standard deviation, the model still gave ca. 90\% correct responses (cf. Table \ref{tab:fracCorrect}). Furthermore, as before, the model was robust to misspecification of the prior uncertainty $\bgs{\std}_\soy$ (see Table \ref{tab:fracCorrect}) suggesting that it is easy to apply the present model to real-world data sets for which the exact amount of noise is unknown.

\section{Discussion}
We have presented a model which switches nonlinear differential equations and have shown that, given observations, filtering using the unscented Kalman filter is sufficient to discriminate which dynamic process produced the observations, in an online fashion and by a simple arg-max readout of the inferred hidden variables. We illustrated the performance of the model using synthetic motion in a plane and realistic point light walker dynamics. By representing a switching process with a Hopfield network with random perturbations, the model achieves a trade-off between stable tracking of a dynamic process in the environment and faithful switching between these processes. Inference performed by the model was robust against noise and model misspecification. 

By using DMPs the model can represent arbitrary dynamic processes. Although we have only shown applications to rhythmic motion, DMPs can also be used to represent goal-directed motion \cite{Ijspeert2003,Schaal2007} which we expect to work equally well with the present switching model. We found that the dynamics implemented by DMPs, i.e., a strongly attracting limit cycle,  is important for successful inference of the switching variables. In preliminary tests we found that when using more flexible dynamic representations, such as standard recurrent neural networks with random connectivity, the model can explain variation due to one of the other dynamical systems by small adaptations of the state of the current dynamical system, as opposed to switching the dynamical system. This is not possible with DMPs, which generate only a single trajectory along a limit cycle. When greater flexibility of the representation is needed, the attractiveness of the limit cycle could be adjusted in the model (parameter $a$ in Section \ref{sec:DMPs}). More importantly, we believe that other parametric models may be used together with the switching dynamics to discriminate potentially more complicated motions, e.g., a hierarchical model which smoothly combines DMPs to represent a long sequence of motions.

The unscented Kalman filter only tracks a single mode of a multimodal posterior. This approximation can impact the discrimination performance of the model. Indeed, we found that the switching of the model during inference on the motion capture data was sometimes delayed (cf. Fig. \ref{fig:walksresults} and Table \ref{tab:fracCorrect}) which may be attributed to this limitation. Nevertheless, we found that the discrimination performance was still robust against random (Gaussian) perturbations which were larger than the actual changes induced by the motion. In particular, the increasing fraction of correct responses with time (cf. Table \ref{tab:fracCorrect}) indicates that, once found, the correct posterior mode was, even under large perturbations, mostly stable in our model.

The performance of the model may be further improved by using inference methods which can approximate the full posterior distribution such as particle filtering. However, this would be bought with an increase in computational demands which might be unnecessary for a given set of dynamic processes, as exemplarily shown here. 


\newpage
\small
\bibliographystyle{ieeetr}
\bibliography{../bitzer_nips2012}

\begin{thebibliography}{10}

\bibitem{Doya2007}
K.~Doya, S.~Ishii, A.~Pouget, and R.~P.~N. Rao, eds., {\em Bayesian Brain}.
\newblock MIT Press, 2007.

\bibitem{Hopfield1984}
J.~J. Hopfield, ``Neurons with graded response have collective computational
  properties like those of two-state neurons.,'' {\em Proc Natl Acad Sci U S
  A}, vol.~81, pp.~3088--3092, May 1984.

\bibitem{Ijspeert2003}
A.~J. Ijspeert, J.~Nakanishi, and S.~Schaal, ``Learning attractor landscapes
  for learning motor primitives,'' in {\em Advances in Neural Information
  Processing Systems 15}, pp.~1523--1530, Cambridge, MA: MIT Press, 2003.

\bibitem{Schaal2007}
S.~Schaal, P.~Mohajerian, and A.~Ijspeert, ``Dynamics systems vs. optimal
  control -- a unifying view,'' in {\em Computational Neuroscience: Theoretical
  Insights into Brain Function} (P.~Cisek, T.~Drew, and J.~F. Kalaska, eds.),
  vol.~165 of {\em Progress in Brain Research}, pp.~425 -- 445, Elsevier, 2007.

\bibitem{Fruhwirth-Schnatter2006}
S.~Fr{\"u}hwirth-Schnatter, {\em Finite Mixture and Markov Switching Models}.
\newblock Springer, 2006.

\bibitem{Chen2000}
R.~Chen and J.~S. Liu, ``Mixture kalman filters,'' {\em Journal of the Royal
  Statistical Society: Series B (Statistical Methodology)}, vol.~62, no.~3,
  pp.~493--508, 2000.

\bibitem{Chen2003}
Z.~Chen, ``Bayesian filtering: From kalman filters to particle filters, and
  beyond,'' tech. rep., Adaptive Syst. Lab., McMaster Univ., Hamilton, Canada,
  2003.

\bibitem{Ghahramani2000}
Z.~Ghahramani and G.~E. Hinton, ``Variational learning for switching
  state-space models.,'' {\em Neural Comput}, vol.~12, pp.~831--864, Apr 2000.

\bibitem{Murphy2002}
K.~P. Murphy, {\em Dynamic Bayesian Networks: Representation, Inference and
  Learning}.
\newblock PhD thesis, University of California, Berkeley, 2002.

\bibitem{Quinn2009}
J.~A. Quinn, C.~K. Williams, and N.~McIntosh, ``Factorial switching linear
  dynamical systems applied to physiological condition monitoring,'' {\em IEEE
  Transactions on Pattern Analysis and Machine Intelligence}, vol.~31,
  pp.~1537--1551, 2009.

\bibitem{Garnett2010}
R.~Garnett, M.~A. Osborne, S.~Reece, A.~Rogers, and S.~J. Roberts, ``Sequential
  bayesian prediction in the presence of changepoints and faults,'' {\em The
  Computer Journal}, vol.~53, no.~9, pp.~1430--1446, 2010.

\bibitem{Saatci2010}
Y.~Saat{\c{c}}i, R.~Turner, and C.~E. Rasmussen, ``Gaussian process change
  point models,'' in {\em Proceedings of the 27th International Conference on
  Machine Learning (ICML-10)} (J.~F{\"u}rnkranz and T.~Joachims, eds.), (Haifa,
  Israel), pp.~927--934, Omnipress, June 2010.

\bibitem{Adams2007}
R.~P. Adams and D.~J. MacKay, ``Bayesian online changepoint detection,'' tech.
  rep., Cambridge, UK, 2007.

\bibitem{Fearnhead2007}
P.~Fearnhead and Z.~Liu, ``On-line inference for multiple changepoint
  problems,'' {\em Journal of the Royal Statistical Society: Series B
  (Statistical Methodology)}, vol.~69, no.~4, pp.~589--605, 2007.

\bibitem{Alvarez2010}
M.~Alvarez, J.~Peters, B.~Schoelkopf, and N.~Lawrence, ``Switched latent force
  models for movement segmentation,'' in {\em Advances in Neural Information
  Processing Systems 23}, pp.~55--63, 2010.

\bibitem{Opper2010}
M.~Opper, A.~Ruttor, and G.~Sanguinetti, ``Approximate inference in continuous
  time gaussian-jump processes,'' in {\em Advances in Neural Information
  Processing Systems 23}, pp.~1831--1839, 2010.

\bibitem{Stimberg2011}
F.~Stimberg, M.~Opper, G.~Sanguinetti, and A.~Ruttor, ``Inference in
  continuous-time change-point models,'' in {\em Advances in Neural Information
  Processing Systems 24}, pp.~2717--2725, 2011.

\bibitem{Stimberg2012}
F.~Stimberg, A.~Ruttor, and M.~Opper, ``Bayesian inference for change points in
  dynamical systems with reusable states - a chinese restaurant process
  approach,'' in {\em JMLR W\&CP 22}, pp.~1117--1124, 2012.

\bibitem{Gams2009}
A.~Gams, A.~Ijspeert, S.~Schaal, and J.~Lenarcic, ``On-line learning and
  modulation of periodic movements with nonlinear dynamical systems,'' {\em
  Autonomous Robots}, vol.~27, pp.~3--23, 2009.
\newblock 10.1007/s10514-009-9118-y.

\bibitem{Jazwinski1970}
A.~H. Jazwinski, {\em Stochastic Processes and Filtering Theory}.
\newblock Academic Press, 1970.

\bibitem{Julier1995}
S.~Julier, J.~Uhlmann, and H.~Durrant-Whyte, ``A new approach for filtering
  nonlinear systems,'' in {\em American Control Conference, 1995. Proceedings
  of the}, vol.~3, pp.~1628 --1632 vol.3, jun 1995.

\bibitem{Wan2001a}
E.~A. Wan and R.~van~der Merwe, ``The unscented kalman filter,'' in {\em Kalman
  Filtering and Neural Networks} (S.~Haykin, ed.), John Wiley \& Sons, Inc.,
  2001.

\bibitem{Doucet2001}
A.~Doucet, N.~de~Freitas, and N.~Gordon, eds., {\em Sequential Monte Carlo
  Methods in Practice}.
\newblock Springer, 2001.

\bibitem{Sarkka2007}
S.~S{\"a}rkk{\"a}, ``On unscented kalman filtering for state estimation of
  continuous-time nonlinear systems,'' {\em Automatic Control, IEEE
  Transactions on}, vol.~52, pp.~1631 --1641, sept. 2007.

\end{thebibliography}

\end{document}